\documentclass[twocolumn, amssymb, amsmath,nobibnotes,nofootinbib, aps, prb,showpacs]{revtex4-2}
\usepackage[dvips]{graphicx}
\usepackage{amsmath}
\usepackage{bm}
\usepackage{color}
\begin{document}
	\title{Evidence of exchange-striction and charge disproportionation in the magneto-electric material  Ni$_3$TeO$_6$}
	\author{Mohamad Numan$^1$, Gangadhar Das$^2$, Md Salman Khan$^3$, Gouranga Manna$^4$, Anupam Banerjee$^1$, Saurav Giri$^1$, Giuliana Aquilanti$^2$, and Subham Majumdar$^1$}
	\email{sspsm2@iacs.res.in}
	\affiliation{$^1$School of Physical Science, Indian Association for the Cultivation of Science, 2A \& B Raja S. C. Mullick Road, Jadavpur, Kolkata 700 032, INDIA}
	\affiliation{$^2$Elettra Sincrotrone Trieste, Strada Statale 14, km 163.5 in AREA Science Park, Basovizza, Trieste 34149, Italy}
	\affiliation{$^3$School of Material Science, Indian Association for the Cultivation of Science, 2A \& B Raja S. C. Mullick Road, Jadavpur, Kolkata 700 032, INDIA}
	\affiliation{$^4$ New Chemistry Unit, Jawaharlal Nehru Centre For Advanced Scientific Research, Rachenahalli Lake Rd, Jakkur, Bengaluru, Karnataka 560064, India}
	
	\begin{abstract}
		The chiral magneto-electric compound Ni$_3$TeO$_6$ is investigated through temperature dependent synchrotron based  powder x-ray diffraction and x-ray absorption spectroscopy between 15 to 300 K. Our work provides a direct  evidence for the exchange-striction in the material around the concomitant onset point of collinear antiferromagnetic and the magneto-electric phases. The x-ray absorption near edge spectra and x-ray photoelectron spectra show that the sample consists of both Ni$^{2+}$ and Ni$^{3+}$ ions in the lattice. The ionic state of Ni is found to be quite robust, and it is largely independent of the preparation route. Additionally, minority  Te$^{4+}$ state is found to coexists with the majority Te$^{6+}$ state, which may arise from the charge disproportionation between Ni and Te ions (Ni$^{2+}$ + Te$^{6+} \iff$ Ni$^{3+}$ + Te$^{4+}$).  The observed mixed valency of Ni is also confirmed by the total paramagnetic moment ($\mu_{eff}$) per Ni atom in the system. This mixed valency in the metal ions and the exchange-striction  may be attributed to the observed magneto-electric effect in the system.
	\end{abstract}	
	\maketitle
	
\section{Introduction}
The intriguing coupling between magnetic and electric orders in solid leads to the fascinating phenomenon of multiferroicity~\cite{Spaldin,Cheong2007,Khomskii,Tokura_2014}. Type II multiferroics, where the electric polarization is driven by the magnetic order,  found to have  large magneto-electric (ME) coupling~\cite{Eerenstein2006}. They often show large ME effect, where the electric polarization can be altered by the magnetic field and vice versa. Type-II multiferroicity in a solid can originate from different underlying mechanisms, such as (i) inverse Dzyaloshinskii-Moriya interaction via antisymmetric spin exchange interaction (spin current mechanism)~\cite{DM_interaction,DM_perov}; (ii) symmetric spin exchange interaction in collinear magnets (Heisenberg exchange striction)~\cite{Exchange_PRL} and (iii) $p$-$d$ hybridization between the metal and the ligand~\cite{p_d_hybridization,p_d,T_Arima}. In certain cases, more than one mechanisms can be present for multiferroicity~\cite{CaMn7O12}.

\par   
In recent times, several 3$d$ transition metal tellurates (M$_3$TeO$_6$, M = Mn, Co, Ni, Cu) having Corundum type structure are found to be relevant materials for multiferroicity. These compounds crystallize  with a chiral structure (space group: \textit{R}3) which  lacks both mirror and inversion symmetry. The interest in these Corundum materials is renewed after the discovery of extremely large ME effect in Ni$_3$TeO$_6$ (NTO). The compound orders antiferromagnetically below $T_N$ = 52 K with a collinear structure~\cite{Zivkovic}. NTO shows  metamagnetic transition under $H$ = 86 kOe at 2 K from a collinear antiferromagnetic  to a incommensurate spiral structure~\cite{INS}. Although, it does not show spontaneous electric polarization ($P$) below $T_N$ under zero magnetic field ($H$ = 0), a large ME effect is observed below $T_N$ under the application of  $H$, and  $P$ attains a value of 3280 $\mu$C/m$^2$ at 2 K  for an applied field of 90 kOe ($P$, $H \parallel c$)~\cite{Oh2014}. Despite the colossal ME effect in NTO is observed beyond a metamagnetic transition at 86 kOe, a significant field induced polarization is present even at lower fields~\cite{Kim,IR_PRB}.

\begin{figure*}
	\centering
	\includegraphics[width = 17 cm]{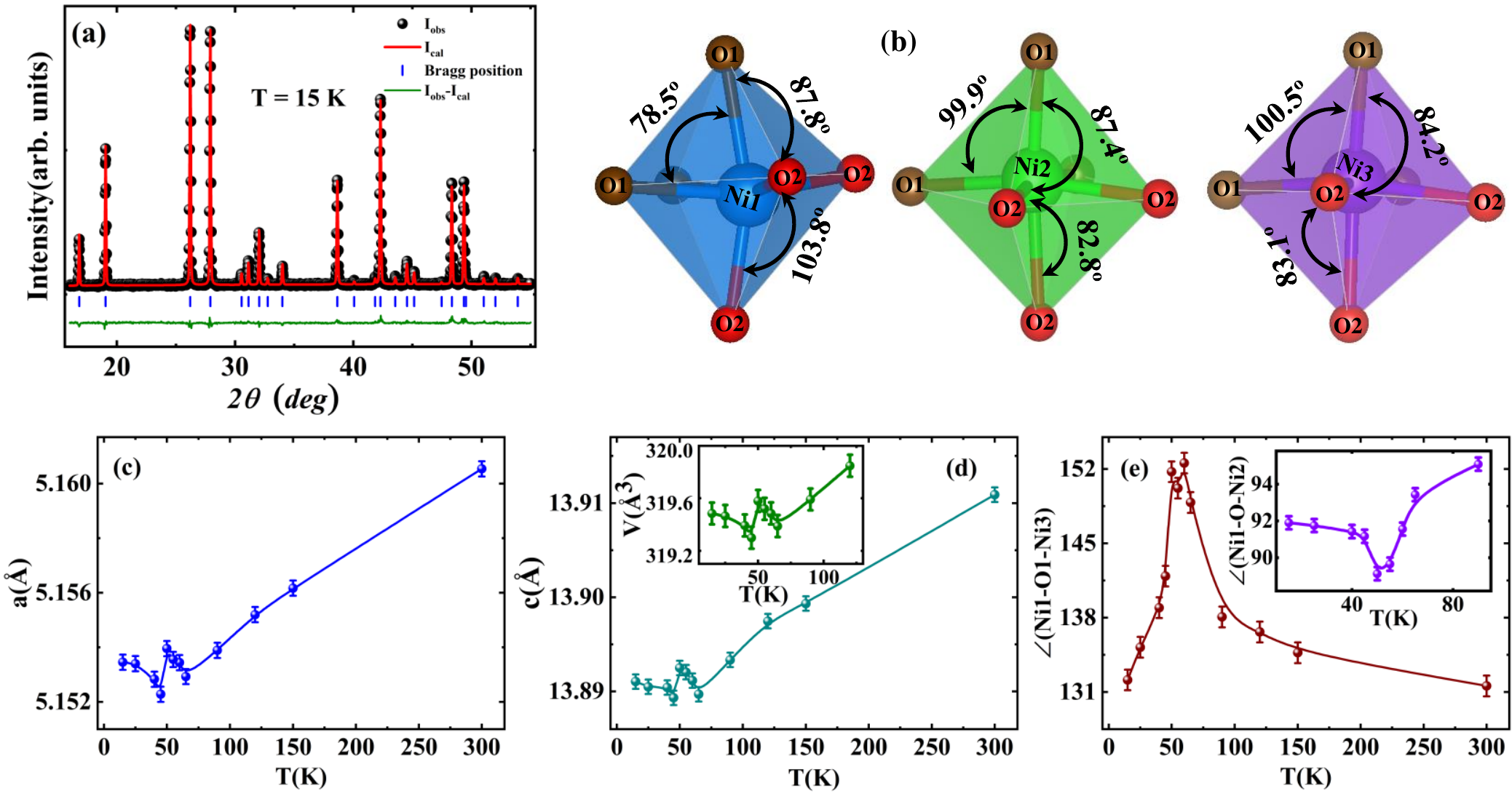}
	\caption{(a) shows the synchrotron x-ray diffraction pattern recorded on NTO-air at 15 K along refinement. (b) Depicts the different bond angle between O-Ni-O inside three different NiO$_6$ ocathedra at three different Ni site. (c) and (d): Thermal variation of lattice parameters ($a$ and $c$) respectively. (e) Illustrates the temperature variation of $\angle{{\rm[Ni(1)-O1-Ni(3)]}}$ bond angle which is associated with the highest magnitude of magnetic exchange interaction path. The insets in (d) and (e) respectively  show the temperature variation of the lattice volume and the bond angle $\angle{{\rm[Ni(1)-O-Ni(2)]}}$ averaged over two oxygen sites.}
	\label{xrd}
\end{figure*}
\par 
In NTO, Ni cations occupy three different crystallographic sites [Ni(1), Ni(2) and Ni(3)], whereas Te cations reside only on a single site. Three NiO$_6$ octahedra [see fig.~\ref{xrd} (b)] and the TeO$_6$ octahedron are found to be face sharing, edge sharing and corner sharing  among each other~\cite{Zivkovic}.  Along the $c$ axis, there are two different pairs of face shared octahedra, one is  between Ni(2)O$_6$ and Ni(3)O$_6$ and the other is between Ni(1)O$_6$ and TeO$_6$. In the $ab$ plane, there are  two different pairs of corner sharing octahedra [Ni(1)O$_6$-Ni(2)O$_6$ and Ni(3)O$_6$-TeO$_6$]. These corner sharing octahedra  form honeycomb like structure in the $ab$ plane with slight offset in the $c$ axis.   

\par
It has been mooted that symmetric exchange-striction can be the possible reason for the ME effect, particularly in the low-field region~\cite{Oh2014}. Exchange striction occurs due to the competition between lattice and magnetic energies in the system. When the system undergoes long range magnetic ordering, the magnetic ions may slightly relocate their position to compensate the additional magnetic energy~\cite{lines}.  Despite the above conjecture of exchange striction, there is hardly any work available in the literature for its  experimental verification. We addressed this problem from the structural point of view using both x-ray diffraction and the x-ray absorption spectroscopy tools. The temperature ($T$) dependent powder  x-ray diffraction (PXRD) shows clear anomaly in the lattice parameters of NTO around $T_N$. The x-ray absorption spectroscopy also indicates noticeable change in the bond length between the metal ion and oxygen (Ni-O) inside the NiO$_6$ octahedra close to $T_N$. The exchange striction is also indirectly evident from the change in  the pre-edge peak area with $T$ as well as the presence of thermal hysteresis in the magnetization versus temperature data. The latter indicates that the magnetic transition is first order in nature.  The charge state of Ni was earlier assumed to be +2. However, our careful investigation of the near edge XAS data indicates that there is a charge disproportionation between Ni and Te sites resulting both Ni$^{2+}$ and Ni$^{3+}$ ions.

\section{Experimental Methods}
Polycrystalline sample of Ni$_3$TeO$_6$ was prepared by standard solid state reaction method using  high purity ($>$99.99\%) starting materials. Stoichiometric amount of NiO and TeO$_2$ were ground and mixed thoroughly, and heated at 750 {\textcelsius} for 15 hours. It was then pressed into pellet and heated at 800 {\textcelsius} for 24 hours. The final heat treatment was done at 830 {\textcelsius}~\cite{R_Sankar,Numan} in air. In addition to this air annealed sample (NTO-air), we also prepared samples which are annealed in vacuum (NTO-vac) and oxygen flow (NTO-oxy)~\cite{NTO-oxy}. Primary measurements were performed on NTO-air, while NTO-vac and NTO-oxy were also investigated to confirm the charge state of Ni through magnetic measurements and photoelectron spectroscopy. The phase purity of the samples was determined using RIGAKU Smartlab (9 KW) XG equipped with Cu K$_\alpha$ (wavelength, $\lambda$ $= 1.5406$ \AA). Beside that, all the $T$-dependent (15-300 K) XRD patterns were obtained using high-flux synchrotron radiation source (energy = $10$ keV) at the Indian Beamline, BL-18B, Photon Factory, Japan. The powder XRD patterns were analyzed using Rietveld technique and the refinement of the crystalline parameter was obtained by using MAUD software package~\cite{Lutterotti}. This confirms that all the samples of Ni3TeO6 are single phase and crystallizes in corundum R3 space group at room temperature. The magnetization ($M$) of the samples was measured using the vibrating sample magnetometer (VSM) module of Quantum Design physical properties measurement system (Dynacool model) as well as Quantum Design SQUID-VSM (MPMS3) in the $T$ range 2-300 K. The temperature dependent (20-300 K) x-ray absorption spectroscopy (XAS) measurements were performed  at the Ni-K absorption edge at the XAFS Beamline of Elettra Sincrotrone Trieste, Italy~\cite{Cicco_2009}. We also recorded the room temperature data at Te-L$_3$ edge spectra. The incident energy was selected using a Si(111) double crystal monochromator. The x-ray absorption near edge structure (XANES) and the extended x-ray absorption fine structure (EXAFS) parts of the acquired XAS data were processed and analyzed using the DEMETER open source package (ATHENA \& ARTEMIS)~\cite{Newville,Ravel}. The x-ray photoemission spectroscopy (XPS)  was performed on a laboratory based  electron spectrometer (Omicron) using  Al K$_{\alpha}$ monochromatic source, and seven-channel Channeltron detector.

\section{Results}
\subsection{Powder x-ray diffraction}
Rietveld refinement of the obtained PXRD patterns at 15 K is shown in Fig.~\ref{xrd}(a), which ensures that the sample retains its room temperature crystal  symmetry at least down to 15 K. As already mentioned there are three different Ni sites, and they form three different distorted NiO$_6$ octahedra. The schematic view of three such  octahedra, as obtained from our PXRD data 15 K, are shown in fig.~\ref{xrd}(b). For an ideal octahedra, the Ni-O-Ni bond angle should be 90$^{\circ}$, and one can see that the actual angles differ considerably from 90$^{\circ}$ in the NiO$_6$ octahedra. The distortion is trigonal type, where there is an elongation of the octahedra along the threefold symmetry axis~\cite{SrTiO3,V2O3,T2g_eg}.

\par
The $T$ variation of the lattice parameters ($a$ and $c$) are shown in figs.~\ref{xrd}(c) and (d). Both $a$ and $c$ decrease with temperature monotonically down to 70 K. A clear anomaly in both the lattice parameters are seen at 60 K, which matches well with the magnetic transition temperature ($T_N \sim$ 54 K) of the sample. This indicates that the system undergoes some structural changes, albeit the lattice symmetry remains the same. Previous density functional theory (DFT) calculation indicates that the strongest Ni-Ni interaction is antiferromagnetic (AFM) type and it occurs via O atom between Ni(1) and Ni(3)~\cite{Wu_DFT}. Considering this fact, we have plotted the $T$ variation  of the bond angle $\angle{{\rm [Ni(1)-O-Ni(3)]}}$ in fig.~\ref{xrd}(e). The data show a prominent peak at around $T_N$ with the peak value of 152$^{\circ}$. This value is closer to 180$^{\circ}$ required for the AFM superexchange in accordence with the Goodenough-Kanamori rule. On the other hand, the bond angle associated with the strongest ferromagnetic (FM) interaction also shows anomaly around $T_N$ with its maximum value of 95$^{\circ}$ [see the inset of fig.~\ref{xrd} (e)], which is favorable for the FM interaction. The observed anomaly at around $T_N$ indicates the presence of strong magneto-structural correlation in the system. It is to be noted that the previous works on the sample also hypothesised the role of exchange striction towards the development of polarization below $T_N$~\cite{Oh2014}. 

\begin{figure}
	\centering
	\includegraphics[width = 8 cm]{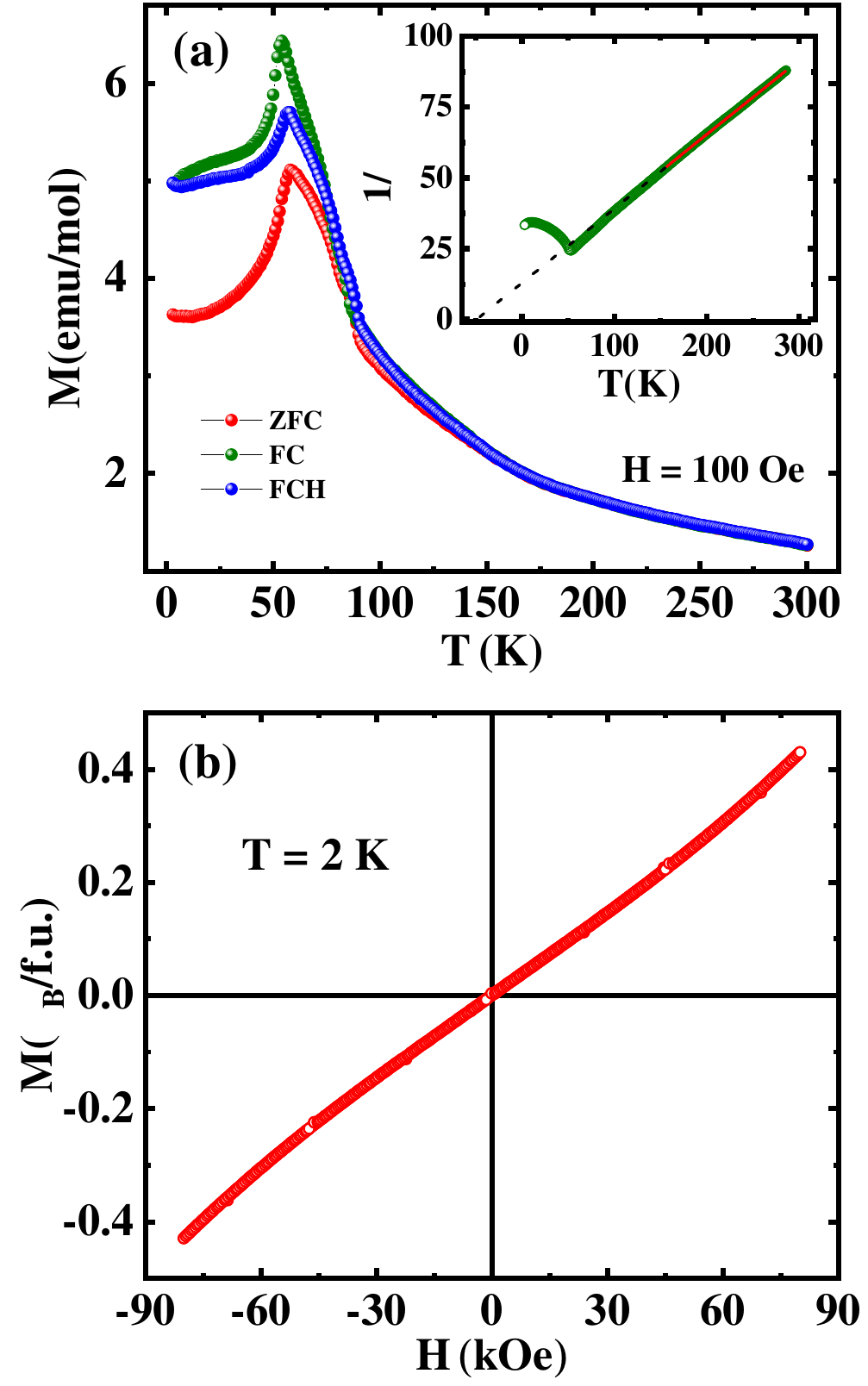}
	\caption{(a) Field-cooling (FC), field-cooled-heating (FCH) and zero-field-cooled-heating (ZFC) magnetization as a function of temperature measured under 100 Oe of field for NTO-air. The inset shows the inverse magnetic susceptibility with a Curie-Weiss fit between 150-300 K. (b) Shows the isothermal magnetization curve at $T$ = 2 K up to a field of $\pm$ 80 kOe.}
	\label{mag}
\end{figure}

\subsection{Magnetization}
Fig.~\ref{mag} shows $M$ as a function of $T$ measured in the field-cooling (FC), field-cooled-heating (FCH) and zero-field-cooled-heating (ZFC) protocols under 100 Oe of applied field ($H$). An AFM like transition is clearly visible at around $T_N$ = 54 K, which match well with the previous reports~\cite{Zivkovic}. The ZFC curve deviates from the FCH data below about 90 K. The FC and FCH data show clear thermal hysteresis over a wide $T$-range around the AFM transition region (4-74 K). This indicates that the transition around $T_N$ is first order in nature. The inverse susceptibility ($1/\chi = H/M$) is found to be linear above about 150 K, and a Curie Weiss fit to the data (see the inset of fig.~\ref{mag}) indicates that the effective moment, $\mu_{eff}$ = 3.32 $\mu_B$/Ni and the paramagnetic Weiss temperature, $\theta_p$ = $-$49 K for NTO-air. For the Ni$^{2+}$ state with spin $S$ = 1, the expected spin-only moment is 2.83 $\mu_B$. The higher value of the effective moment indicates either orbital contribution, or due to a different charge state of Ni. The $\mu_{eff}$ of NTO-vac and NTO-oxy are found to be 3.32  $\mu_B$/Ni and 3.35  $\mu_B$/Ni respectively, indicating that  the moment value of NTO does not depend upon the preparation condition. We recorded the isothermal $M$ versus $H$ data, which are depicted in  fig.~\ref{mag} (b). The curve is quasilinear in the low field region (below 50 kOe), and a clear upward curvature is seen at higher $H$. One can relate this curvature to  the metamagnetism observed in the sample reported earlier~\cite{Kim}.

\begin{figure}
	\centering
	\includegraphics[width = 8 cm]{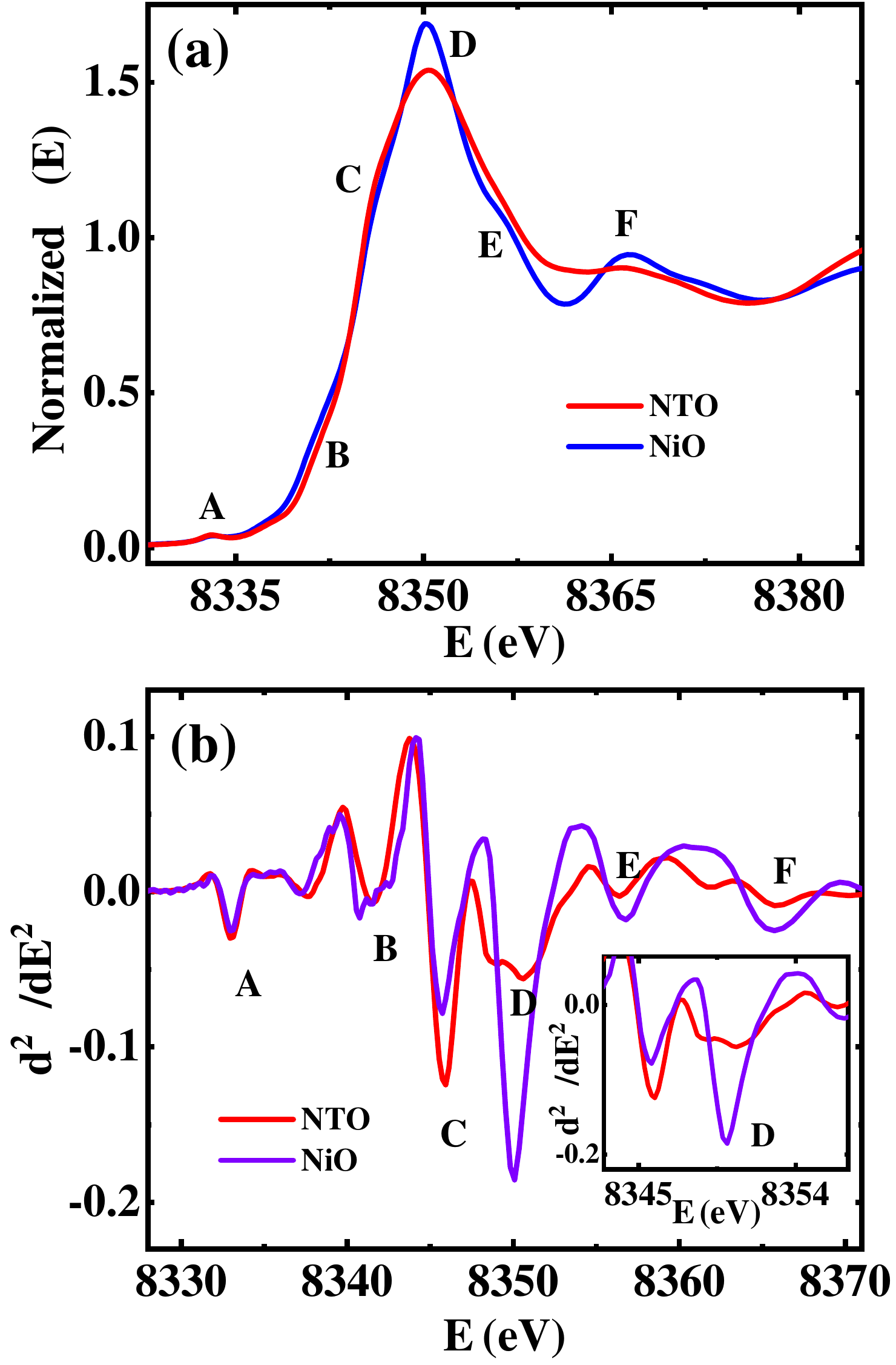}
	\caption{(a) shows the normalized Ni-K absorption edge measured at 300 K for NTO-air and the standard NiO. (b) Second derivative of the absorption coefficient as a function of incident energy. The inset shows an enlarged view of data near the main peak. }
	\label{xanes}
\end{figure}

\subsection{X-ray absorption near edge structure}
To maintain the stoichiometric formula of Ni$_3$TeO$_6$, it is reasonable to assume that  Ni and Te are in the  2+ and 6+ states respectively.  However, it is known that Ni and Te can assume multiple charge states which can create mixed valency  at the  Ni and Te sites. In addition, oxygen off-stoichiometry  can also create mixed valency of the cations~\cite{Off_stoichi_2020,JPCL_2016}. Such mixed valency can greatly alter the strength of the  magnetic interactions between Ni ions.

\subsubsection{Ni-K edge}

To reaffirm the charge states of the cations, we have analyzed the XANES spectra at the Ni-K edge as shown in fig.~\ref{xanes}. The obtained spectra were aligned using standard Ni metal foil data in the reference channel. The XANES spectra at the Ni-K edge shows few distinct features, which are designated as A-F in Fig.~\ref{xanes}. These features are originating from different kinds of electronic transitions from 1$s$ level to  higher energy states of the Ni atom in the distorted octahedral environment of oxygen. We also recorded the Ni-K edge spectra of NiO in the same set up along with the sample.

{\bf Pre-edge:} A small hump like feature designated as A in fig.~\ref{xanes}(a)  around 8333 eV is known as the pre-edge. This is a dipole forbidden transition from 1$s$  to  3$d$ electronic states of Ni, which is weakly allowed via quadrupole transition. A prominent pre-edge can occur due to the admixture of 3$d$ level with 4$p$ level of the metal provided the active metal site is in non-centrosymmetric environment~\cite{XANES_manganites,Sanchez}. A pre-edge region can also occur due to the dipole allowed transition from 1$s$ to the hybridized state of Ni-3$d$ and O-2$p$ levels.

\par
\begin{figure}
	\centering
	\includegraphics[width = 8 cm]{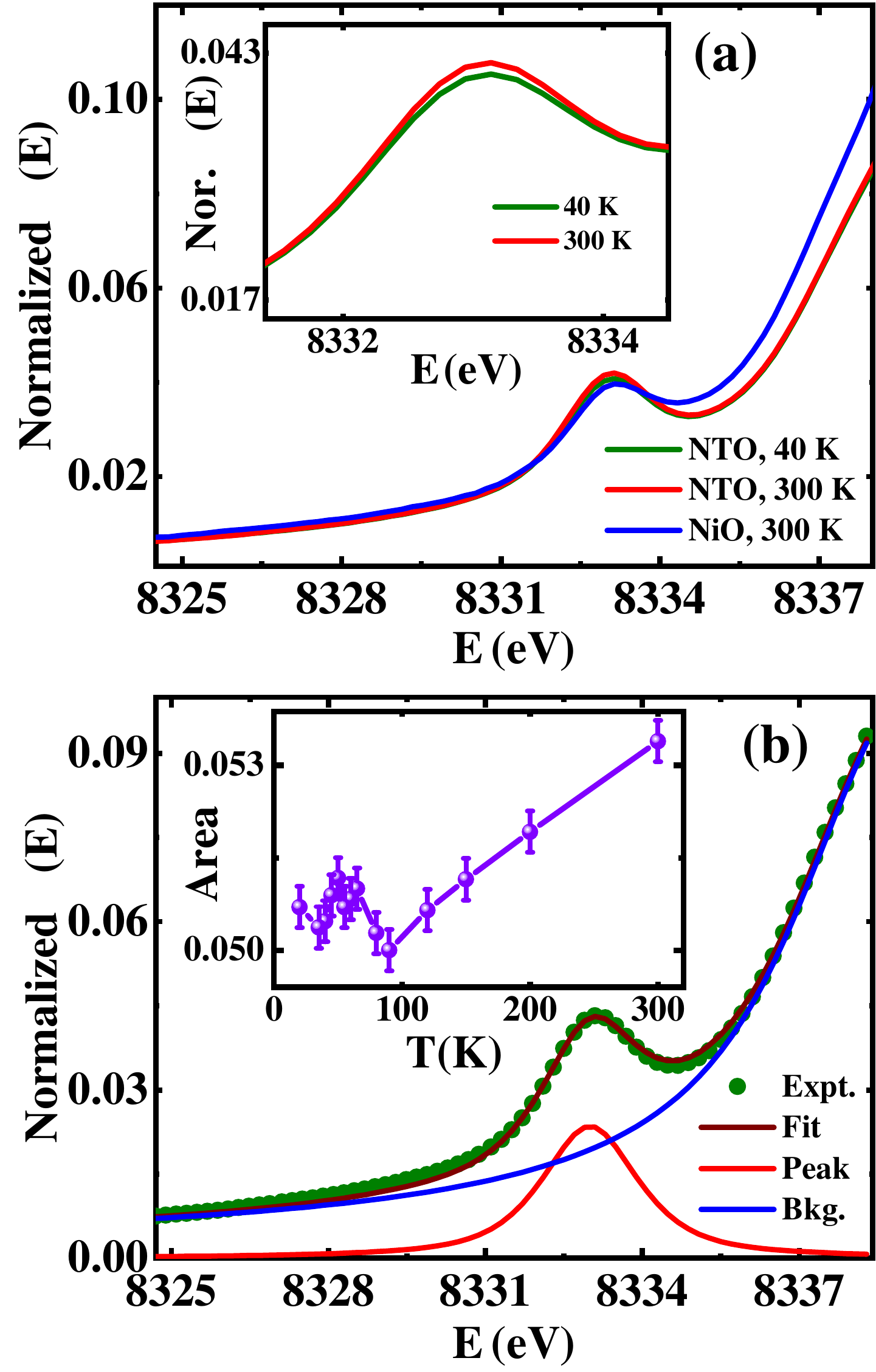}
	\caption{(a) shows the normalized pre-edge of the Ni-K absorption data at 40 and 300 K for NTO-air. The NiO pre-edge is also shown for the comparison. The inset shows an enlarged view of the normalized pre-edge measured at both the temperatures. (b) shows the fitting of the 300 K pre-edge data using arc tangent function (blue line as background)  and a pseudo-voigt function (red line). The inset shows  the intensity (area under the curve) of the pre-edge with temperature.}
	\label{preedge}
\end{figure}

We have fitted the pre-edge using the combination of an  arc-tangent (background) and a pseudo-voigt (pre-edge peak) functions as illustrated in fig.~\ref{preedge}(b) for the 300 K data. The pseudo-voigt peak position is found to be at 8333 eV and this position is independent of temperature.  The area under the pre-edge peak (pseudo-voigt) decreases almost linearly with decreasing $T$ [see inset of fig.~\ref{preedge} (b)] down to about 80 K. Similar linear variation of the pre-edge peak area was observed in other transition metals, and it is possibly related to the dynamic displacement of the absorbing atoms~\cite{Durmeyer_2010,Yoshiasa}. Interestingly, the area under pre-edge shows a non-monotonic variation below 80 K. The pre-edge intensity shows a distinct peak, which matches well with the $T_N$ of the sample. Notably, we find that the pre-edge peak area of NiO is about 23\% less than that of the NTO at 300 K [see table~\ref{tab_preedge}].

\begin{table}[h!]
	\centering
	\begin{tabular}{c@{\hskip 1cm} c @{\hskip 1cm} c}
		\hline \hline
		Sample & Temperature(K) & Area under the curve\\
		\hline
		Ni$_3$TeO$_6$ & 300 & 0.0528(9)\\
		&120 & 0.0501(5)\\
		&35 & 0.0498(7)\\
		\hline
		NiO & 300 & 0.0407(2)\\
		\hline \hline
	\end{tabular}
	\caption{Temperature variation of the area under the fitted (pseudo-voigt) curve of the pre-edge}
	\label{tab_preedge}
\end{table}

{\bf Main absorption peak:} The main absorption peak in the XANES spectra originates from the electronic transition from the occupied inner level to unoccupied outer levels. In this case the absorption of x-ray energy is occurring because of the dipole allowed transition of electrons  from 1$s$ to 4$p$ levels of Ni. Further increase in the x-ray energy leads to transition to the continuum. In fig.~\ref{xanes} (a), the XANES  spectra of NTO along with NiO are plotted. The positions of the XANES rising line in NTO is shifted towards the higher energy as compared to NiO.  The first derivative of the absorption coefficients ($d\mu(E)/dE$) for NTO at the onset of rising edge is found to shifted ($\sim$ 0.2 eV) towards higher $E$ than that of NiO. This implies that Ni in NTO is not completely in the 2+ state, and some fraction of Ni ions are in the higher oxidation state. 

\begin{figure}
	\centering
	\includegraphics[width = 8 cm]{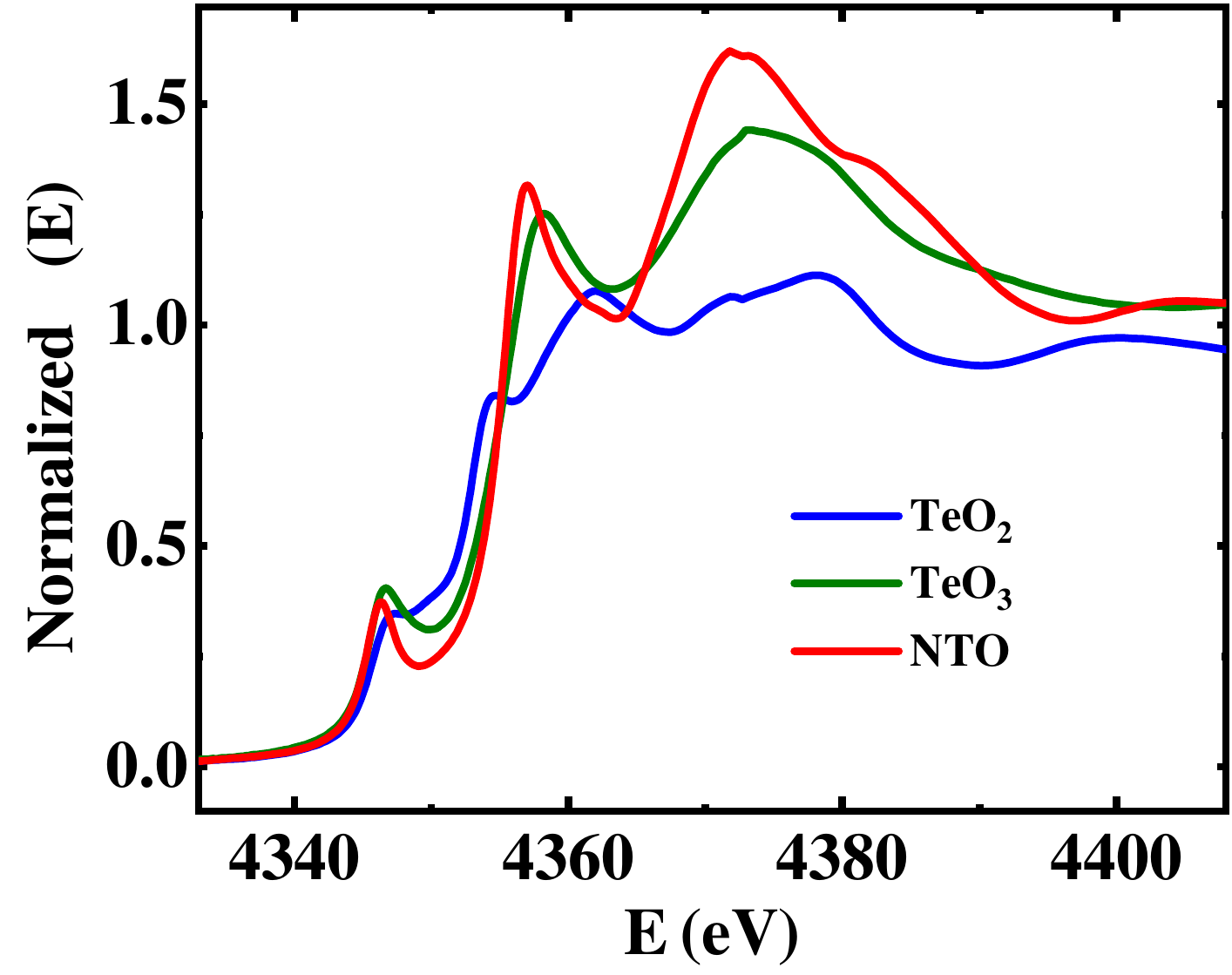}
	\caption{Shows the Te-L$_3$ absorption edge measured at 300 K for NTO-air as well as the standard TeO$_2$(Te$^{4+}$) and TeO$_3$(Te$^{6+}$).}
	\label{Te_xanes}
\end{figure}
	
\par
The shift in the energy position is also clearly visible in the plot of second derivatives of the absorption coefficient ($d^2\mu(E)/dE^2$)  [fig.~\ref{xanes}(b)]. It is to be noted that the nature of the plot is significantly different at the point D (main peak). $d^2\mu(E)/dE^2$ is quite sharp at D for NiO, while it is broad and split into two for NTO-air. Interestingly, for the Li doped NiO samples, where Ni is mixed valent (2+ and 3+), identical feature in $d^2\mu(E)/dE^2$ is observed~\cite{Jacabson}. This also an another indication of the mixed valent state of NTO. Below the main peak, we observe few anomalies at the points B and C, which are also reflected in the $d^2\mu(E)/dE^2$ plot. Such features are often observed in the XANES data of metal oxides, and generally attributed to the charge transfer from metal to ligand~\cite{Aich_2018}.       

\begin{figure*}
	\centering
	\includegraphics[width = 14 cm]{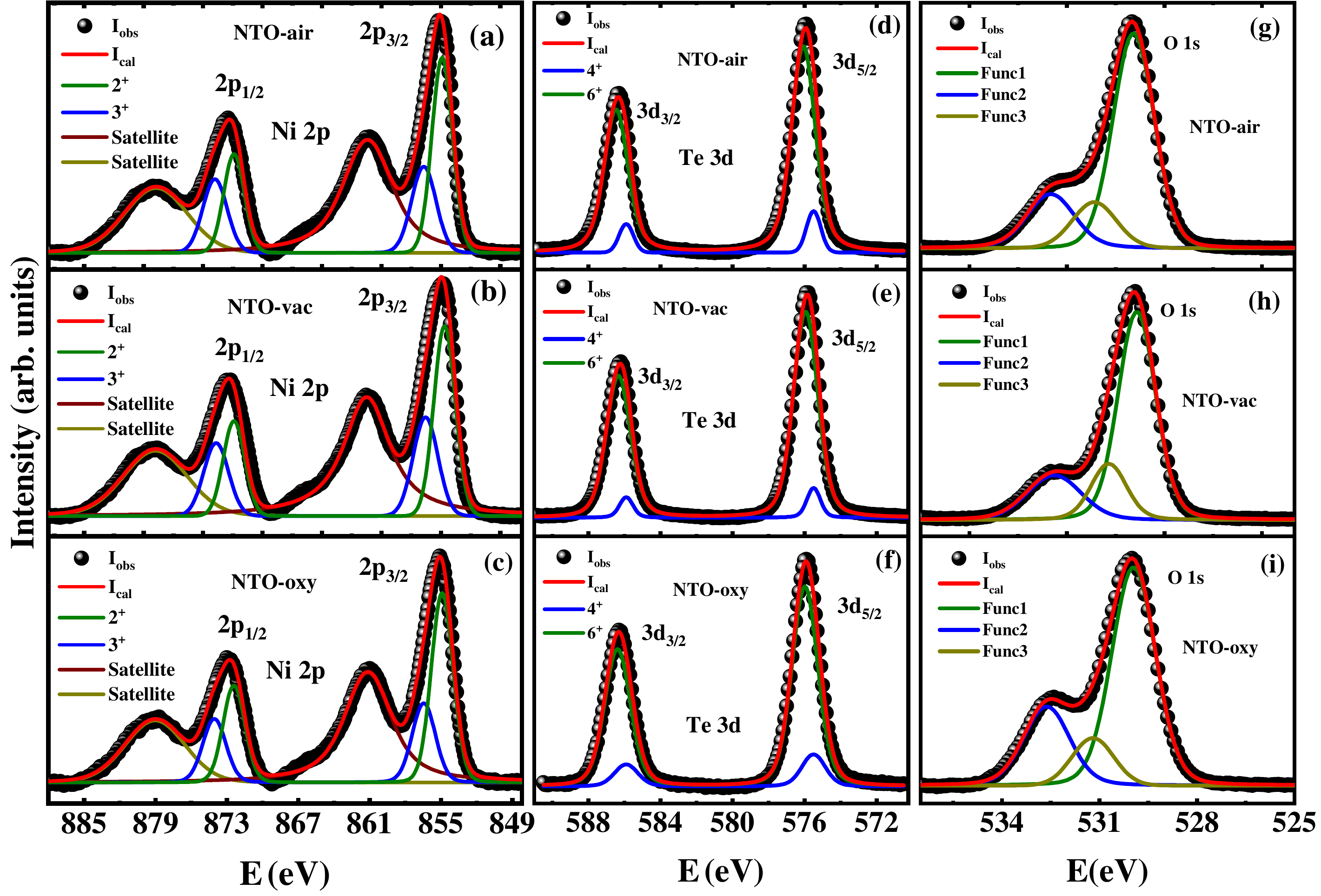}
	\caption{(a)-(c) show the Ni-2$p$ x-ray photoemission peaks for the sample annealed in air, vacuum and oxygen respectively. The Te-3$d$ peaks for different samples are shown in (d)-(f), while the O-1$s$ peaks are depicted in (g)-(i).}
	\label{xps}
\end{figure*}

\subsubsection{Te-L$_3$ edge}
We also recorded the spectra of the Te-L$_3$ edge as shown in fig.~\ref{Te_xanes} along with the standard samples of TeO$_2$ (Te in 4+) and TeO$_3$ (Te in 6+). Evidently, the L$_3$ rising edge of NTO lies between that of TeO$_2$ and TeO$_3$, although it is closer to the TeO$_3$ sample~\cite{Te_H_Singh,Te_PRB,Te_Glass}. The first derivative of absorption coefficient at the rising edge also shows that the peak for NTO resides somewhat close to TeO$_3$ albeit a at slightly lower energy ($\sim$ 0.5 eV). It indicates that majority of the Te ions are in the 6+ state, although certain fractions are also in the 4+ state.   
 
\subsection {X-ray photoelectron spectroscopy}
In order to investigate  the oxygen off-stoichiometry towards the mixed valency of Ni, we studied three different batches of Ni$_3$TeO$_{6}$ samples, namely NTO-air, NTO-vac and NTO-oxy. The obtained data were calibrated using the standard carbon 1$s$ peak located at 285.4 eV. Such vacuum and oxygen  annealings are likely to change  oxygen stoichiometry in Ni$_3$TeO$_{6}$. We  recorded x-ray photoemission spectra of O-1$s$, Ni-2$p$ and Te-3$d$ levels for all three samples [see figs.~\ref{xps} (a)-(i)]. The Ni-2$p$ spectra show two distinct set of peaks (2$p_{1/2}$ + satellite and 2$p_{3/2}$ + satellite) originating from the spin orbit interaction. The Ni-2$p_{3/2}$ shows a main peak at 855.1 eV along with a satellite line at 861.2 eV. For Ni$^{2+}$ state (as in NiO), the binding energy (BE) of  2$p_{3/2}$ level is about 854.5 eV ~\cite{XPS_1,XPS_6}. This indicates that the BE of 2$p_{3/2}$ in NTO is slightly higher than the pure 2+ state. This is possibly due to the presence of Ni in higher oxidation state, which is also evident from our XANES study. We failed to fit the  Ni-2$p_{3/2}$ line with a single Lorentz convoluted Gaussian line. Nevertheless, two Lorentz convoluted Gaussians provide good fit to the data [fig.~\ref{xps} (a)], and they correspond to two charge states of Ni, namely Ni$^{2+}$ and Ni$^{3+}$. Interestingly, there is practically no difference in peak position and the profile of the Ni-2$p_{3/2}$ lines in NTO-air and NTO-vac and NTO-oxy indicating that annealing in  different environments hardly changes the Ni charge state [figs.~\ref{xps} (a)-(c)]. 

\par
We also recorded the Te-3$d$ line as depicted in fig.~\ref{xps} (d)-(f) for the NTO-air, NTO-vac and NTO-oxy samples respectively. Similar to Ni peaks, the Te-3$d$ peaks are found to be identical in all three samples. The 3$d$ line of Te is split into 3$d_{3/2}$ and 3$d_{5/2}$ due to spin-orbit coupling. The 3$d_{5/2}$ line of NTO occurs at 575.9 eV, which is slightly lower than the BE of 576.4 eV reported for Te$^{6+}$ state (as in TeO$_3$)~\cite{TeO3_JPCM,TeO3_AIP}. Such observation further indicates that Te is also in a mixed valent state similar to Ni. The O-1$s$ line of the NTO sample is observed around 530 eV, and there exists a shoulder like feature around 532.5 eV [fig.~\ref{xps} (g)]. The peak can be fitted with three Lorentz convoluted Gaussian singlets. The O-1$s$ peaks of NTO-vac and NTO-oxy samples are found to be almost identical with that of NTO-air [fig.~\ref{xps} (g)-(i)].

\begin{figure*}
	\centering
	\includegraphics[width = 14 cm]{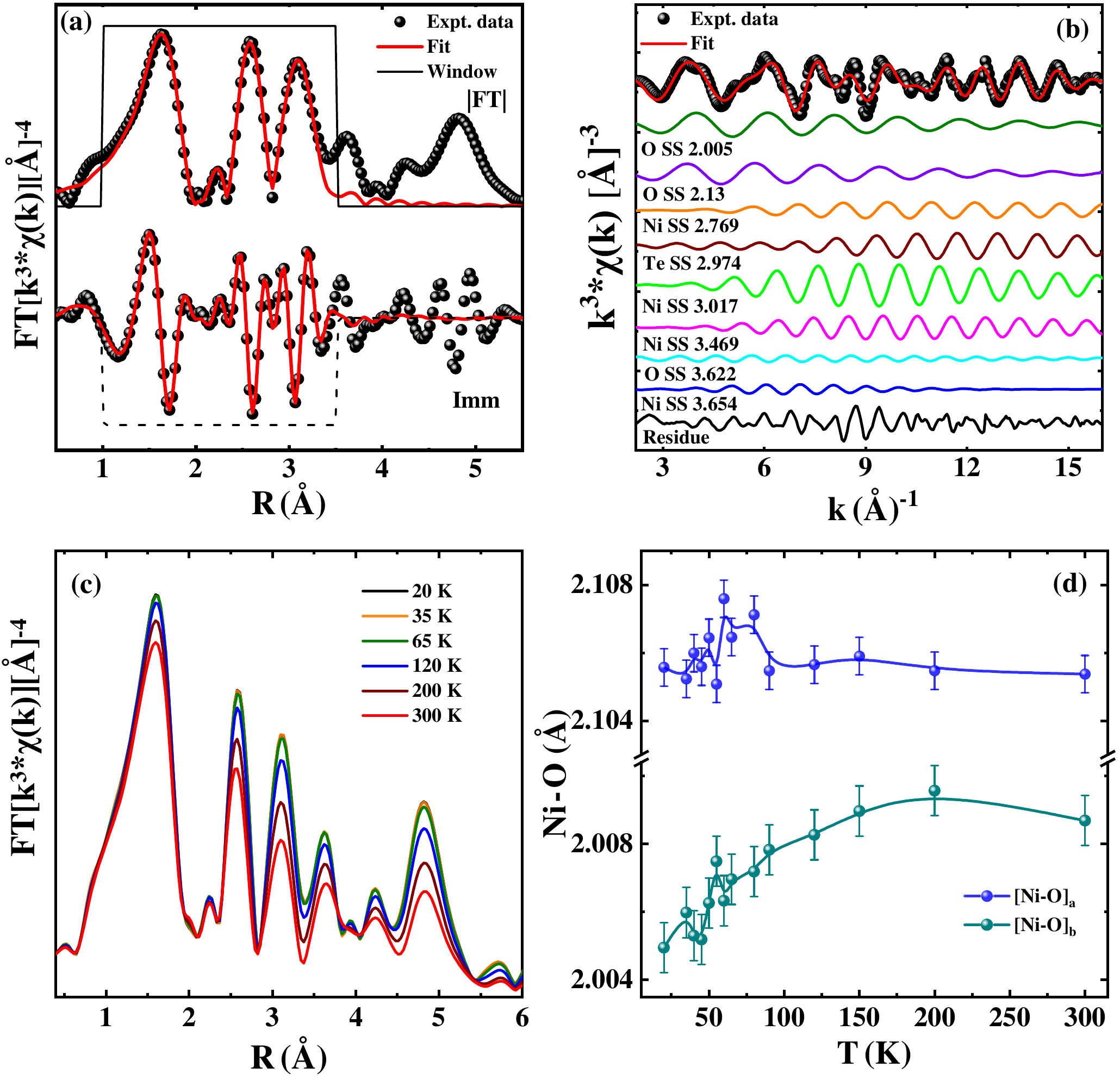}
	\caption{(a) The Fourier transforms of the respective experimental data (black circles) and the theoretical fitted curve (solid red line); the magnitude ($|$FT$|$) and the imaginary parts ($I_{mm}$) are indicated, and they are  vertically shifted for clarity. (b) Ni-K edge $k^3$ weighted experimental EXAFS data (shaded black circles) at $T$ = 300 K and the corresponding best fit (red solid line). (c) shows the $T$ variation of the radial distribution function with $R$. The variation of the Ni-O bond lengths inside the Ni-O$_6$ octahedra with $T$ are depicted in (d). The two different sets of oxygen inside three different Ni-O$_6$ octahedra are designated as [Ni-O]$_a$ and [Ni-O]$_b$ respectively. }
	\label{exafs}
\end{figure*}

\subsection{Extended x-ray absorption fine structure}
To have a precise information about the temperature evolution of the local structure of NTO, we have analyzed the EXAFS part of the absorption data recorded between 300 and 20 K. Fig~\ref{exafs}(a) shows the imaginary part and the modulus of the Fourier transform ($|$FT$|$) of the EXAFS signal of the data at 300 K compared with the best fit up to 3.5 \AA.  We have used FEFF6 to  calculate the phase shifts and effective scattering amplitudes of the atoms~\cite{Newville_FEFF}. Since, Ni has three different (in-equivalent) crystallographic sites, we have used an aggregate FEFF calculation. The scattering paths within 0.15 \AA~were merged together, weighted by the fractional population of the site in the unit cell~\cite{Ravel_FEEF}. The $k^3$ weighted experimental spectra with the best fit along with different scattering paths are depicted in fig.~\ref{exafs}(b) and the partial contributions from different shells are shown below for better visualization. It is to be noted that, the scattering paths at 3.622  \AA~and 3.654 \AA~are considered for a better fit to the data below 3.5 \AA.

\par
The temperature variations of the $|$FT$|$ of the $k^3$ weighted experimental EXAFS data are shown in Fig.~\ref{exafs}(c). There is a systematic increase in the amplitude of the $|$FT$|$ data with the lowering of $T$, which is likely to be connected with the decrease of the Debye-Waller factor with decreasing $T$. Additionally, there is a weak shift of the  peak around 2.6~\AA~to higher values of $R$ with lowering of $T$. From our EXAFS analysis, we conclude that this peak primarily originate from adjacent Ni and Te atoms. Therefore, the anomaly may be attributed to the change in Ni-Ni and Ni-Te bond lengths with $T$. 
\par 
Fig.~\ref{exafs} (d) shows the $T$ variation of Ni-O bond length for two different oxygen sites. We find an  anomaly in the data around $T_N$, which supports similar feature observed  in the lattice parameter [see fig.~\ref{xrd}(c)-(e)].

\begin{figure*}
	\centering
	\includegraphics[width = 15 cm]{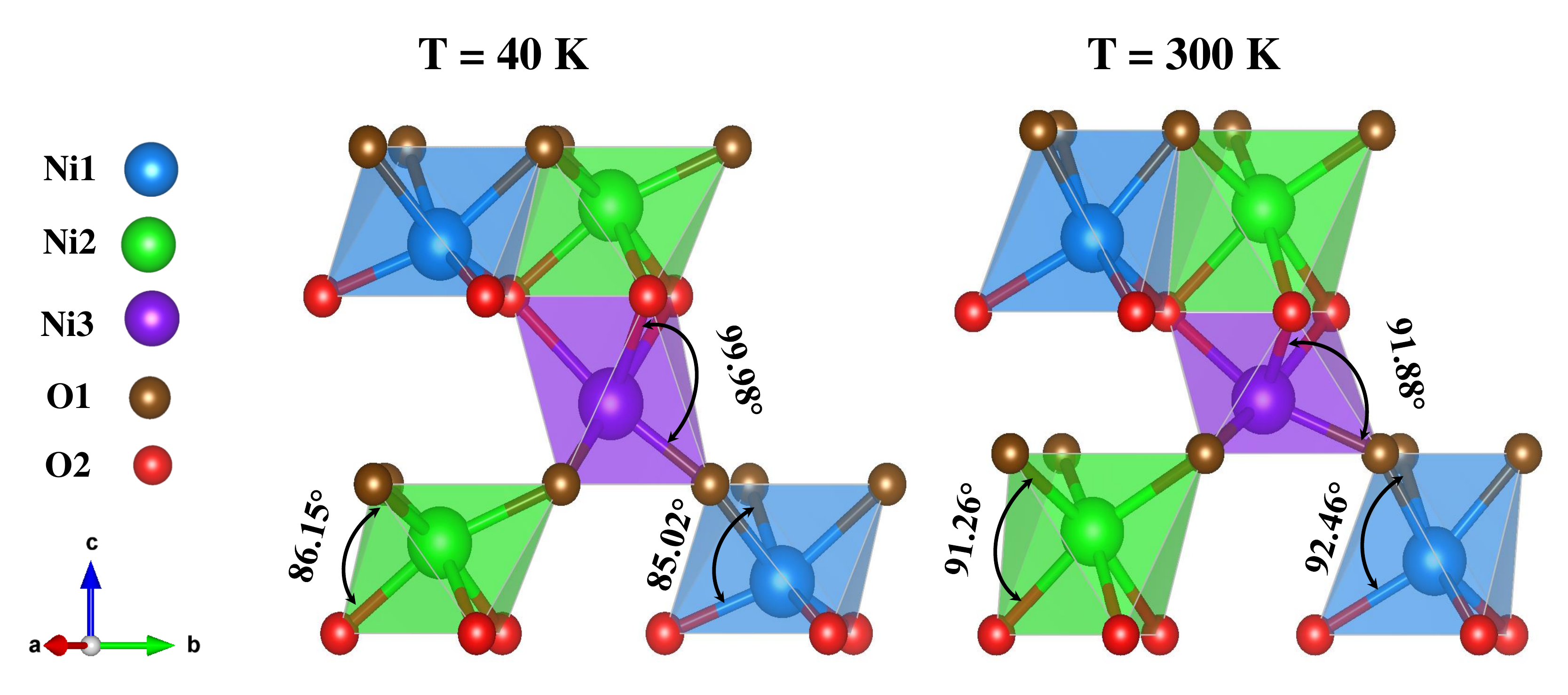}
	\caption{Perspective views of the NiO$_6$ octahedra at 40 K and 300 K.}
	\label{distort}
\end{figure*}

\section{Discussions} 	
 Despite having non-centrosymmetric crystal structure, NTO does not show switchable spontaneous polarization down to 2 K. However, below the long range antiferromagnetic transition at $T_N$ = 52 K, the sample shows non-switchable electric polarization ($P$) along with colossal magneto-electric effect. The compound has been referred as a pyroelectric material. 
 
 The present work brings out several new interesting facts on the compound.
 
 \begin{itemize}
 \item {Our magnetic measurement indicates clear evidence for the thermal hysteresis around $T_N$. The  hysteresis spans a wide temperature range. Previous works identified the magnetic transition to be second order in nature~\cite{Oh2014}. However, the presence of thermal hysteresis clearly rules out the possibility of a continuous transition in NTO.}
 	
 \item {Our PXRD indicates that the crystal symmetry remains unchanged down to 15 K. Nevertheless, an anomaly is observed in the lattice parameters around $T_N$. Such anomaly supports the conjecture of exchange striction mechanism for the observed magneto-electric effect. From our EXAFS data, we also find similar anomaly close to $T_N$ in the Ni-O bond length. Such observation provides a direct proof for the exchange striction in the system.}
 	
 \item {As a consequence of the lattice anomaly,  sharp discontinuity is also present in the   lattice volume [see the inset of fig.~\ref{xrd}(a)]. Since volume is the first derivative of Gibb's free energy $V = \left( \frac{\partial G }{\partial P}\right )_T$, a discontinuity in the lattice volume indicates the first order nature of the transition. This is at par with the observed thermal hysteresis in our magnetization data.}
 
 \item{$\angle$[O-Ni-O] in the NiO$_6$ octahedra are close to 90$^{\circ}$ at 300 K. However, a larger deviation from 90$^{\circ}$ is observed in all $\angle$[O-Ni-O] at lower temperature as shown in fig.~\ref{distort}. This is due to the trigonal distortion in all three Ni-octahedra associated with the exchange-striction.}
 
 \item {An interesting outcome of our XANES data is regarding the charge state of Ni. In previous works till now, the oxidation state of Ni was considered to be in the 2+ state. However, the main-edge of the XANES data of NTO is shifted to higher energy values as compared to NiO. This indicates that Ni is not in the pure 2+ state and at least some fraction of Ni ions attains higher charge state. It is more likely that some Ni ions of the present sample  are actually in the 3+ state.  In our previous XPS study on NTO, we found evidence of both Ni$^{2+}$ and Ni$^{3+}$ ions~\cite{Numan}. Interestingly, from the Te-3$d$ XPS data and the Te-L$_3$ XANES spectra, we found the existence of majority Te$^{6+}$ along with some fraction of Te$^{4+}$ ions. This might happen due to  charge disproportionation between Ni and Te ions, Ni$^{2+}$ + Te$^{6+} \iff$ Ni$^{3+}$ + Te$^{4+}$. Similar charge disproportionation is often found in transition metal based double perovskites~\cite{abhisek,charge}.}
 
 \item{The Ni pre-edge in NTO is found to be more prominent than the NiO. This may be due to the distorted NiO$_6$ octahedra leading to greater overlap of O-2$p$ and Ni-3$d$ states. Most interestingly, the intensity of the pre-edge peak shows a distinct anomaly at the AFM ordering temperature. Such anomaly can be attributed to the exchange-striction observed in our PXRD data [see figs.~\ref{xrd} (c-d)]. The Ni-O-Ni AFM super-exchange mechanism can also have some effect on the  transition probability associated with the pre-edge leading to the anomaly in the pre-edge peak area at $T_N$.}
 
 \item {The value of the effective paramagnet moment also support the mixed valency in Ni. In the literature, there is a wide values of the reported Ni moment for NTO~\cite{Zivkovic,R_Sankar,Nanograin}.  In the most cases, the value of  $\mu_{eff}$ is higher than the spin only value (2.83 $\mu_B$) of Ni$^{2+}$ moment.  We carefully addressed the issue, and the value of $\mu_{eff}$ is found to be 3.32 $\mu_B$/Ni for NTO-air. It is to be noted that the value of $\mu_{eff}$ for NTO-air is found to be almost same as NTO-vac and NTO-oxy. This value may either arise from the spin orbit coupling (SOC) or due to the presence of Ni$^{3+}$ ions. Recent DFT calculations indicate that the introduction of SOC term in the Hamiltonian gives non-collinear magnetic structure of the sample~\cite{Wu_DFT}, which is not realized in practice. So, the origin of higher moment can be ascribed to  Ni$^{3+}$ ions, which is also supported by our XANES study.}
 
 \item{Our XPS study on three different batches of samples (air, vacuum and oxygen annealed respectively) does not show any marked difference in the O, Ni and Te peaks with annealing procedure. This indicates that the compound is quite stable and its oxygen stoichiometry does not change much with the vacuum annealing. The signature of mixed valency in Ni and Te is present in the XPS data of air, vacuum and oxygen annealed samples.} 
\end{itemize}  

 \par
 Our work brings out two important outcomes of the studied NTO sample. Firstly, though the compound retains its crystal symmetry unchanged down to low temperature, there is a definite change in the lattice parameter at the magnetic ordering temperature. Being a collinear antiferromagnet,  $H$ should have minimum effect on the  electric properties as long as it is below the metamagnetic field. In fact the conical spiral spin structure develops in the compound above $H$ = 86 kOe~\cite{INS}, though ME effect is present at much lower $H$. Our observation of lattice anomaly at $T_N$ indicates the presence of magneto-structural instability. The lattice anomaly due to exchange striction occurs from the competition between magnetic and lattice energies, and the system stabilizes with an optimized separation of magnetic ions. The application of $H$ can  disrupt this equilibrium leading to further change in the ionic separation, which in its turn can induce the ME effect. 
 
\par  
 Secondly, the NTO compounds are found to be mixed valent in both Ni and Te. There are several examples in the literature~\cite{CaBaFe4O7_ME,CaMn7O12_Jhuma,RKKY_ME}, where multiferroicity or ME effect is related to the mixed valency of the magnetic ions. For example, the mixed valent state of Mn and Fe are important for the magneto-electric effect in CaMn$_7$O$_{12}$ and CaBaFe$_4$O$_7$ respectively. The orthrhombic mixed valent manganites RMn$_2$O$_5$ (R = rare-earth) also show giant ME effects, which is thought to arise from the  parity-breaking Mn$^{3+}$-Mn$^{4+}$ isotropic exchange interaction~\cite{CaMn7O12_Jhuma,CaBaFe4O7_ME,TbMn2O5,YMn2O5,RMn2O5}. In case of CaMn$_7$O$_{12}$, CaBaFe$_4$O$_7$, and RMn$_2$O$_5$, the mixed valency of the metal ions is site specific. However, for NTO it is difficult to say whether Ni$^{3+}$ or Ni$^{2+}$ ions occupy definite crystallographic site or they are randomly distributed in the lattice. Nevertheless, the mixed valency in NTO can be a possible route to the large ME effect in the compound.

\par
In conclusion, we have performed a comprehensive analysis of the structural details of the magneto-electric compound Ni$_3$TeO$_6$ through detailed PXRD, XPS and XAS studies. An iso-structural instability is seen at the N\'eel temperature, which can be assigned to exchange-striction effect. The observed structural anomaly can be seen from the local probe such as EXAFS, and it is also evident in the overall change from our PXRD analysis. Our work also points out the mixed valence states of cations, which is presumably related to the charge disproportionation between Ni and Te ions. Such charge disproportionation can be instrumental in the observed magnetic and electric anomalies in Ni$_3$TeO$_6$.  

\section*{Acknowledgments}
MN would like to thank CSIR, India for his research fellowship [File No. 09/080(1131)/2019-EMR-I]. Department of Science and Technology (India) is acknowledged for financial support (KEK Proposal No. 2021-IB-32), and Saha Institute of Nuclear Physics and Jawaharlal Nehru Centre for Advanced Scientific Research, India for facilitating the experiments at the Indian Beam Line, Photon Factory, KEK, Japan. We also acknowledge Elettra Sincrotrone Trieste for providing access to its synchrotron radiation facilities with XAFS beamline (Proposal No. 20210469).

\bibliography{ref_NTO}

\begin{thebibliography}{55}
\bibitem [{NTO()}]{NTO-oxy}%
  \BibitemOpen
  \href@noop {} {}\bibinfo {note} {For the preparation of NTO-oxy, TeO$_3$ was
  prepared by heating H$_6$TeO$_6$ at 350 {\textcelsius} for 4 h. Then the
  stoichiometric amount of NiO and the as prepared TeO$_3$ were mixed and
  grounded thoroughly and heated at 750 {\textcelsius} for 10 h under constant
  oxygen flow. After that, the same heat treatment was followed as NTO-air
  except that this time the heating was done under continuous oxygen
  flow}\BibitemShut {NoStop}%
\end{thebibliography}
\bibliographystyle{apsrev4-2}
	
\end{document}